\RequirePackage{fix-cm}
\documentclass[twocolumn]{svjour3}          
\smartqed  
\usepackage{graphicx}
\usepackage{bm}
\usepackage{amsmath}
\usepackage{tabularx}
\usepackage{longtable}
\usepackage{minibox}
\usepackage{setspace}

\usepackage[utf8]{inputenc}
\usepackage{times}
\usepackage{graphicx}
\usepackage{latexsym}
\usepackage{dsfont}
 \usepackage{amsmath} 
 \usepackage{amsfonts}
 \usepackage{bm}
\usepackage{amssymb}
\usepackage{graphicx}
\usepackage{tabularx}
\usepackage{float}
\usepackage{times}
\usepackage{helvet}
\usepackage{courier}
\usepackage{url,graphicx,times}
\usepackage{subfigure}
\usepackage{tabularx}
\usepackage{amsmath}
\usepackage{amssymb}
\usepackage{multirow}
\usepackage{booktabs}
\usepackage{longtable}
\usepackage{enumerate}
\usepackage{enumitem}
\usepackage{mdwlist}
\usepackage{marvosym}
\usepackage{bm}
\usepackage{dsfont}
\usepackage{comment}
\usepackage{lipsum}
\usepackage[ruled,vlined]{algorithm2e}
\usepackage{soul}
\usepackage{footmisc}
\usepackage{minibox}
\usepackage{framed}
\graphicspath{{figures/}}
\usepackage{tikz}
\usetikzlibrary{fit,positioning}
\usepackage{setspace}
\usepackage{fixmath}
\usepackage{algpseudocode}
\usepackage{mathrsfs}
\usepackage{color}
\begin{document}

\title{NEMR: Network Embedding on Metric of Relation
}
\subtitle{}


\author{Luodi  Xie$^1$, Hong Shen$^1$, Jiaxin Ren$^1$
}


\institute{F. Author \at
              first address \\
              Tel.: +123-45-678910\\
              Fax: +123-45-678910\\
              \email{xield@gmail2.edu.cn}           
           \and
}

\date{Received: date / Accepted: date}

\maketitle

\begin{abstract}

Network embedding maps the nodes of a given network into a low-dimensional space such that the semantic similarities among the nodes can be effectively inferred. Most existing approaches use inner-product of node embeddings to measure the similarity between nodes leading to the fact that they lack the capacity to capture complex relationships among nodes.  Besides, they take the path in the network just as structural auxiliary information when inferring node embeddings, while paths in the network are formed with rich user informations which are semantically relevant and cannot be ignored.
\par In this paper, We propose a novel method called Network Embedding on the Metric of Relation, abbreviated as NEMR, which can learn the embeddings of nodes in a relational metric space efficiently. First, our NEMR models the relationships among nodes in a metric space with deep learning methods including variational inference that maps the relationship of nodes to a gaussian distribution so as to capture the uncertainties. Secondly, our NEMR considers not only the equivalence of multiple-paths but also the natural order of a single-path when inferring embeddings of nodes, which makes NEMR can capture the multiple relationships among nodes since multiple paths
contain rich user information, e.g., age, hobby and profession. Experimental results on several public datasets show that the NEMR outperforms the state-of-the-art methods on relevant inference tasks including link prediction and node classification.
\end{abstract}

\section{Introduction}
\label{intro}
With the idea of mapping the entities into a low-dimensional semantic space, network embedding aims at inferring representations
of entities i.e., nodes and attributes, in the networks, such that the similarities among entities can be measured based on their embeddings. Network embedding has attracted considerable interest in recent years because of its wide applications in the fields of social networks, e.g.,link prediction~\cite{li2018streaming}, node classification~\cite{li2017attributed,perozzi2014deepwalk} and community detection~\cite{wang2017community} etc. 

\par Based on deep learning techniques, numerous network embedding algorithms have been proposed~\cite{li2017attributed,zhang2017user,tang2015pte,perozzi2014deepwalk,huang2017label,cao2015grarep} in recent years.  Inspired by  Skip-Gram model~\cite{mikolov2013efficient}, DeepWalk~\cite{perozzi2014deepwalk}, Node2Vec~\cite{grover2016node2vec} and extract a number of sequential nodes in the network by random walk, make the nodes in a sequence metaphorically equivalent to words in language models, employ Skip-Gram to obtain the embeddings of nodes and then calculate the similarity of nodes with the inner-product of embeddings. 
LINE~\cite{tang2015line} captures the network structure by preserving first - and second-order proximities among nodes.

\par Some other works infer embeddings for networks with rich auxiliary information, such as labels~\cite{huang2017label,chen2017pne:}, textual information~\cite{liu2018content} and topological structure~\cite{kipf2017semi-supervised}.  Cao et al.~\cite{cao2015grarep} proposes an algorithm to take the global and local structure informations into account so as to well capture the relationships between remote nodes.
 Kipf et al.~\cite{kipf2017semi-supervised} learn node embeddings through aggregating features from neighborhoods. Huang et al.~\cite{huang2017accelerated} employ the graph Laplacian technique to learn the joint embeddings from the topological structure of the network. 

\par Existing methods enjoy impressive success, but they still suffer from a number of main drawbacks:
\begin{itemize}
\item They use the inner-product of node embeddings to measure the similarity between nodes, which leads to the lack of capacity to capture complex relationships of nodes, e.g., the relationship among neighborhoods~\cite{hsieh2017collaborative}, and the uncertainty of relationship.
\item They take the path between a node pair as structural auxiliary information when inferring node embeddings, which ignore the rich semantic informations hidden in the paths. Each path represents one social relationship between a node-pair. In contrast, a node-pair which has only one path between them tends to have a simple social relationship.
\end{itemize}

\par To address the aforementioned defects, in this paper, we propose a novel method, Network Embedding on Metric of Relation, abbreviated as NEMR, which can learn the embeddings of nodes in a relational metric space effectively. The proposed NEMR models the relationships between nodes in a semantic metric space with deep learning methods including variational inference that maps the relationships of nodes to gaussian distributions so as to capture the uncertainties of the relationships. NEMR infers embeddings of nodes with both the equivalence of multiple-paths and the natural order of a single-path, which can capture more complex multiple relationships among nodes.

\par How to learn the embeddings of nodes in a relational metric space effectively is still a challenge. First of all, in the vector space, the affinities between entities is rather complex and can not be measured simply by using inner-product metric, like what previous work do. Furthermore, due to the fact that nodes may play different roles under different conditions, the relationships among nodes tend to be uncertain.
Rather than using traditional inner-product metric to model the affinities between entities~\cite{perozzi2014deepwalk,grover2016node2vec,tang2015line,pan2016tri,qiu2018network}, a novel probabilistic metric method needs to be proposed, with considering the uncertainty of metric space. Besides, in the real world, between two nodes, there are often more than one path, each of which contains rich user information and represents one social relationship between node pair, which brings more challenges to modeling the relationships between entities and learn the embeddings of entities in the semantic metric space correspondingly.

\par To address these challenges, we utilize deep learning methods including variational inference to model the relationships among nodes in a metric space. As VAE (Variational Auto-encoder) is a powerful generative model to obtain the probability distributions of the latent variable, it is
applicable to model uncertainty. Thus, we integrate variational inference into modeling affinities between entities to capture the uncertainty of the relationship between nodes in the metric space. Considering the case of multiple paths existing between two nodes, each of which takes along rich information of users, we take the multiple paths into account when inferring embedding to capture multiple social relations
between node pair.

\par The main contributions of our approach can be summarized as follows:

\begin{itemize}
\item We propose a novel unsupervised embedding model, called NEMR,  which models the relationship between any two nodes in a semantic metric space  to capture the complex relationships between nodes.    
\item We propose the equivalence of multiple-paths between node-pairs which is firstly defined in our paper. Multiple-paths between node-pairs are rich in social relationships and our NEMR can capture them better. 
\item We propose the natural order of a single-path between two nodes which is also firstly defined in our paper. Keeping the nodes on the single-path in natural order is essentially a good way to describe the simple relationship between them.
\item To verify the effectiveness of NEMR algorithm, we conduct extensive experiments on several public datasets. Experimental results show that NEMR can learn better embeddings of nodes and outperforms the state-of-the-art methods.
\end{itemize}
\par The remainder of this paper is structured as follows. In Section 2, we briefly review the related work on network embedding and metric learning. Three definitions and two problem formulations are described in Section 3. Section 4 introduces the proposed algorithm NEMR in detail. And in Section 5, we compare our NEMR model with the baselines on several public datasets. Finally, we conclude our work in Section 6.

\section{Related Work}
\subsection{Metric Learning}
Metric learning, also known as similarity learning, was first proposed in~\cite{xing2002distance}. The basic idea of metric learning is to learn a particular metric function according to different tasks and the learned function can map the similarity between two samples into a metric space. Metric learning has been applied into many areas, e.g., image classification~\cite{qian2015fine}, text retrieval~\cite{lebanon2006metric} and text classification~\cite{wang2005learning}, due to its powerful expansibility. Recently, Collaborative Metric Learning (CML)~\cite{hsieh2017collaborative} shows that metric learning can satisfy the crucial triangle inequality which allows it to capture fine-grained relationships between two samples. Metric learning is also applied into network embedding~\cite{cheng2019network,chen2018pme:}.  DML-NRL\cite{cheng2019network} can capture the information of both network structure and labels of nodes through metric learning. Chen et al.\cite{chen2018pme:} learn a distance metric to preserve both the first-order and the second-order proximities in heterogeneous network. But to the best of our knowledge, to capture complex relationships between nodes, few previous work considers modeling the relationships among nodes in a semantic metric space.
\subsection{Variational Auto-encoders}
AE (Auto-encoder)~\cite{bourlard1988auto-association} is an unsupervised model which can obtain the latent representations of input data. AE consists of two computational neural networks: one names encoder, aiming at encoding the input data $x$ into latent representations $z$, and another one names decoder, aiming at reconstructing the input data $x^{\prime}$ back from the latent encoded representations. When $x$ and $x^{\prime}$ are sufficiently close, we assume that $z$ is the  latent representation of $x$. VAE (Variational Auto-encoder)~\cite{kingma2013auto-encoding} is an extension of AE, which assumes that the output latent variable $z$ obeys a prior distribution, e.g., gaussian distribution. VAE is a generative model, and its goal is to obtain the probability distribution of the latent variable $z$ under the observation sample $x$. By using Stochastic Gradient Descend~\cite{kingma2015adam:} and reparameterization tricks~\cite{kingma2013auto-encoding}, the variational Evidence lower BOund(ELBO) can be optimized to approximate the log-likelihood iteratively. The powerful ability to acquire latent representations of VAE allows it applied into many tasks. Many variations of VAE have been proposed~\cite{jiang2016variational,kingma2014semi-supervised,kipf2016variational}, and they have been applied into various tasks such as semi-supervised classification~\cite{kingma2014semi-supervised}, clustering~\cite{creswell2019denoising} and image generation~\cite{dosovitskiy2016generating}.

\subsection{Network Embedding}
Network embedding maps the entities of networks into a low-dimensional semantic space. Existing models of network embedding are concerning about the technology of mapping or dimension reduction. An efficient way for dimension reduction is matrix factorization. The more accurately a matrix factorization include higher-order information, the better effect it will have, along with higher computational complexity. Based on matrix factorization, numerous methods are proposed~\cite{cao2015grarep,yang2017fast,yang2015network}. Yang et al.~\cite{yang2015network} prove that the essence of DeepWalk~\cite{perozzi2014deepwalk} is matrix factorization and they make use of the text feature information in the process of  Matrix factorization.
GraRep~\cite{cao2015grarep} takes the global and local structure information into account, and this method can capture the relationships between remote nodes. NEU~\cite{yang2017fast} summarizes some embedding methods which can be regarded as matrix factorization. 

Another efficient way for dimension reduction is deep learning. DeepWalk~\cite{perozzi2014deepwalk} and Node2vec~\cite{grover2016node2vec} translate the network structure into statements in corpus by random walk. SDNE~\cite{wang2016structural} adopt a deep auto-encoder to preserve both the first-order and second-order proximities.  Some embedding models of heterogeneous networks are also proposed~\cite{tang2015pte,liu2017semantic}. The graph variational auto-encoder (GAE)~\cite{kipf2016variational} is an unsupervised manner  to learning on graph-structured data based on the variational auto-encoder(VAE).
TRIDNR\cite{pan2016tri} is an extension of DeepWalk model~\cite{perozzi2014deepwalk}, which considers network structures, the contexts of nodes and the labels of nodes when inferring node embeddings.
LINE~\cite{tang2015line} uses first-order and the second-order proximities, and trains the model via negative sampling.

Some approaches of graph convolutional networks ~\cite{kipf2017semi-supervised,hamilton2017inductive} have been proposed.  Kipf et al.\cite{kipf2017semi-supervised}  uses an efficient layer-wise propagation rule that is based on a first-order approximation of spectral convolutions on graphs. GraphSAGE\cite{hamilton2017inductive} is an inductive model suitable for large-scale network, which can quickly generate embedding for new nodes without additional training process. Methods based on knowledge graph has also been proposed in recent years~\cite{bordes2013translating,wang2014knowledge}. The basic idea of these embedding models is regarding every relation as translation in embedding space.

Few approach considers the idea of learning an embedding with a probability distribution. \cite{vilnis2014word}  is the first one to learn Gaussian word embeddings to capture uncertainty. \\Graph2Gauss \cite{bojchevski2017deep} embeds each node as a gaussian distribution, and learns network structure via a novel unsupervised personalized ranking formulation. However, these embedding methods cannot capture the uncertainty of relationships between nodes.

\section{NOTATIONS AND PROBLEM FORMULATION}
Before we describe our method, we first introduce some notations and terminology, and then formally define the problem to be addressed.

\subsection{Notations and Definitions}
\textbf{DEFINITION 1 (Social Network)}. Let $\mathcal{G}=(\bm{\mathcal{V}}, \bm{\mathcal{E}}, \mathbf{A})$ be a graph (directed/undirected), where $\bm{\mathcal{V}}$ = $\{ v_i \}_{i=1}^N$ denotes the set of nodes, $\bm{\mathcal{E}}  = \{ e_{i,j}\}_{i,j=1}^{E}$ denotes the set of edges, $\mathbf{A}$ is the adjacency matrix for the network, $N$ is the total number of nodes, and $E$ is the total number of edges. We let $\textbf{A}_{i,j}=1$ denote that there is an edge between node $v_i$ and node $v_j$, and let $\textbf{A}_{i,j}=0$ represent there is no edge between node $v_i$ and node $v_j$.

\textbf{DEFINITION 2 (Social Relationship)} In a social network $\mathcal{G}=(\bm{\mathcal{V}}, \bm{\mathcal{E}}, \mathbf{A})$ (directed/undirected),  social relationships refer to the connections formed by different users due to a series of factors such as work, interests, blood ties, etc,  which can be measured by the similarities among nodes in metric space.  A social relationship between node $v_i$ and node $v_j$, i.e., $Sim_{(i,j)}$ is the similarity between $v_i$ and $v_j$.

\textbf{DEFINITION 3 (Social Relationship Path)} In a social network $\mathcal{G}=(\bm{\mathcal{V}}, \bm{\mathcal{E}}, \mathbf{A})$ (directed/undirected), a social relationship path, can be represented as a set of variable-length node sequences.  As shown in Fig.1, there are three paths between node $v_1$ and node $v_4$, i.e., $P_{1}= \{ v_1, v_2, v_3, \\v_4 \}$, $P_{2}= \{ v_1, v_4 \}$, $P_{3}= \{ v_1, v_5, v_4 \}$, and thus there are three relationships correspondingly. Users on the path $P_{1}= \{ v_1, v_2, v_3, v_4 \}$ are a small group with the common interest on basketball, u sers on the path $P_{2}= \{ v_1, v_4 \}$ are a small group with the common interest on football, and Users on the path $P_{3}= \{ v_1, v_5, v_4 \}$ are a small group with the common interest on badminton.

\begin{figure}[!t]
\center
\includegraphics[width=0.7\columnwidth]{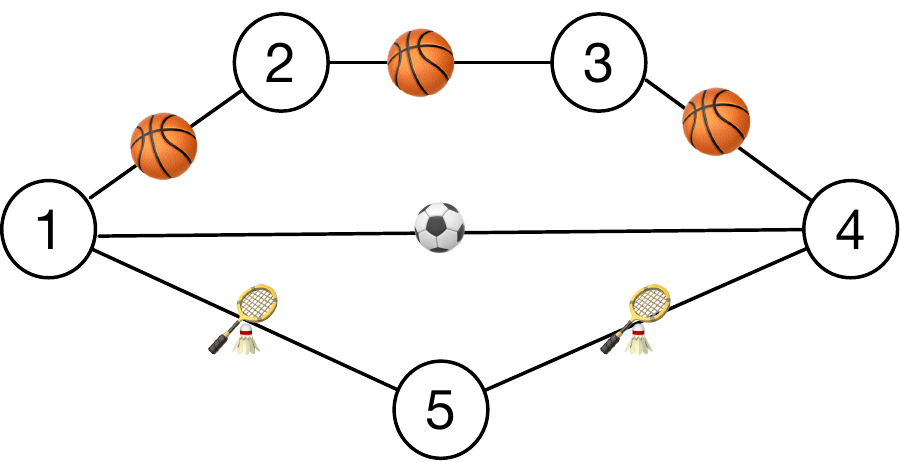}
\caption{\textbf{Definition 4: Multiple-paths between a  node-pair, i.e., node $v_1$ and $v_4$,  due to different social relationships. There are three paths between nodes $v_1$ and $v_4$, each of which represents one relationship, i.e., football, basketball and badminton. In metric space, these three paths are equivalent, what means:  $Sim_{(v_1,v_4)} $=$ Sim_{(v_1,v_2)}$ + $ Sim_{(v_2,v_3)} $ + $Sim_{(v_3,v_4)}  $= $  Sim_{(v_1,v_5)} $ + $ Sim_{(v_5,v_4)} $ . }}
\label{fig:figure1}
\end{figure}

Despite there are multiple paths(relationships) between node pairs, they all indicate the connectivity of node pairs in a global view. To measure multiple relationships between node pair, we map the relationships into metric space, assuming that the multiple paths between node pair follows the equivalence of multiple-paths, which is described in Definition 4.

\par \textbf{DEFINITION 4 (The equivalence of multiple-paths).}  For a node pair $i$ and $j$, $\bm{\mathcal{S}}_{i,j}$ is the set of paths between $i$ and $j$, $P \in \bm{\mathcal{S}}_{i,j}$ represents a path between node $i$ and $j$, which can be represented by the node sequence $\{ v_1, v_2, v_3 ... v_{t-1}, \\v_{t} \}$, and $R_{P}$ is the total relationship of path $P$, i.e., $R_{P}= Sim_{(v_1,v_2)}+Sim_{(v_2,v_3)}+...+Sim_{(v_{t-1},v_{t})}$. The equivalence of multiple-paths means that, for arbitrary path $P, Q \in \bm{\mathcal{S}}_{i,j}$, $R_{P} = R_{Q}$. A scenario of Definition 4 is shown in Fig.1.

There are a few node pairs with only one path between them, the relationship between these nodes is relatively simple. For these node pairs, the proposed equivalence between multiple-paths can not capture the relationship between them. As shown in Fig.2, a solid line indicates that there is a relationship between two nodes, and a dotted line means there is no direct relationship between two nodes. $P_1 = \{ v_5, v_4, v_1,\\ v_2, v_3\}$ and $P_2 = \{ v_7, v_6, v_1, v_8, v_9\}$ represent two single-paths in the graph, node-pairs on these two paths have just one social relationship(violin or bowling). To capture the relationship between nodes on a single-path, we assume that the relationship between them follows a natural order, which is described in definition 6.

\textbf{DEFINITION 5 (Direct distance).} In a social network $\mathcal{G}=(\bm{\mathcal{V}}, \bm{\mathcal{E}}, \mathbf{A})$ (directed/undirected), for node $i$ and $j$, direct distance between $i$ and $j$ in metric space, i.e., $r'_{i,j}$, is the similarity between $i$ and $j$, i.e., $Sim_{(i,j)}$, despite there is not an direct relationship between $i$ and $j$.

\textbf{DEFINITION 6 (The natural order of a single-path).}  Let $\bm{\mathcal{T}}$ be the set of single-paths in the network, $P_{i,j} \in \bm{\mathcal{T}}$ represents the only one path between node $i$ and $j$, which can be represented by a node sequence $\{ v_1, v_2, v_3 ...... v_{t-1}, \\v_{t} \} $. The natural order of a single path means that, for each node pair $i$ and $j$ on $P$, if node $i$ and $j$ don't have a direct relationship between them, then $r'_{i,j} > R_{i,j}$, where $r'_{i,j}$ is the direct distance between $i$ and $j$ in metric space, $R_{i,j}$ is the total relationship between node $i$ and $j$ on the single path, i.e., $R_{i,j}=Sim_{(v_1,v_2)}+Sim_{(v_2,v_3)}+...+Sim_{(v_{t-1},v_{t})}$. A scenario of Definition 6 is shown in Fig.2.

For ease of reference, in Table 1, we list the basic notations used in this paper.
\begin{figure}[!t]
\center
\includegraphics[width=0.7\columnwidth]{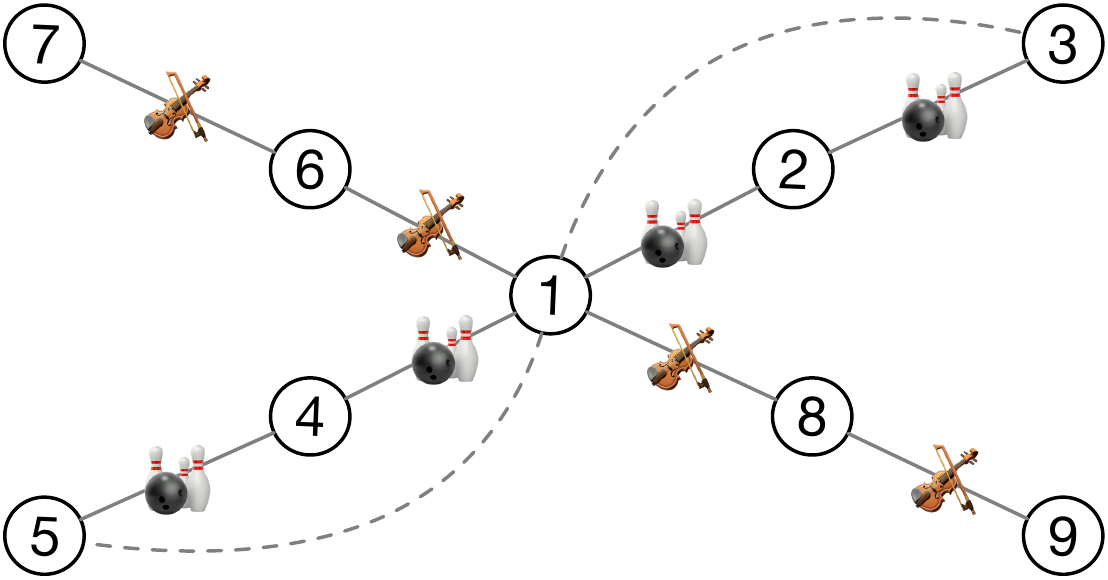}
\caption{\textbf{ Definition 6: A single-path between node-pairs, i.e., node $v_1$ and $v_5$. A solid line indicates that two nodes have a social relationship, and a dotted line indicates not. There is only one path between nodes $v_1$ and $v_5$, and $r'_{1,5}$ represents the direct distance between node $v_1$ and $v_5$. In metric space, nodes on a single path maintain a natural order, what means: $r'_{1,5}$ $>$ $Sim(1,4)$ + $Sim(4,5)$ }}.
\label{fig:figure2}
\end{figure}

\subsection{Problem Definition}
With the terminologies described above, the two problems to be solved in our paper are defined as follows:
\par \textbf{Problem 1. Node Embedding.} 
Given a undirected graph $\mathcal{G}$, the goal is to learn the latent representation of nodes with the mapping function $\Xi_{1}$:

\begin{align}
\mathcal{G}  \stackrel{\Xi_{1}} {\longrightarrow}   \mathbf{\Phi}^{\mathcal{V}}
\end{align}

such that structure information of the network can be preserved as much as possible by $\mathbf{\Phi}^{\mathcal{V}} \in \mathcal{R}^{N \times K}$. $\mathbf{\Phi}_{i}^{\mathcal{V}}$ being the embedding of the $i-th$ node to be inferred.
\par \textbf{Problem 2. Learning Representation of relationship in Metric Space .} Given the latent embedding matrix for all nodes $\mathbf{\Phi}^{\mathcal{V} }$, the goal is to learn a metric function $\Xi_{2}$  that satisfies the following formulation in an unsupervised way:
\begin{equation}
\mathbf{\Phi}^{\mathcal{V}} \stackrel{\Xi_{2}}{\longrightarrow} \mathbf{Z}^{\mathcal{E} }
\end{equation}
for node $i$ and $j$, $\mathbf{Z}_{i,j}^{\mathcal{E} } $ is the latent representation of relationships to be inferred with $\mathbf{\Phi}_{i}^{\mathcal{V}}$ and $\mathbf{\Phi}_{j}^{\mathcal{V}}$.

\begin{table}
  \caption{Main notation used across the whole paper.}
  \label{notation}
 \resizebox{0.5\textwidth}{!}{
  \begin{tabular}{@{~}lll}
    \toprule
    \textbf{Notation} & \textbf{Desription}\\
    \midrule
 $\bm{\mathcal{G}}$             &              a directed graph                           \\
$\bm{\mathcal{V}} $               &              set of nodes			 	\\
$\bm{\mathcal{E}} $                &             set of edges				\\
$\bm{\mathcal{S}}$ 			 &		set of multiple-paths		\\
$\bm{\mathcal{T}}$			&		set of single-paths			\\
$N$                                          &            number of nodes				 \\
$E$                                          &            number of edges				 \\		
$K$                                          &            embedding dimension of nodes and relationships \\                 		
$\mathbf{A} \in \mathcal{R}^{N \times N}$	                                 &		adjacency matrix of nodes					\\
$\mathbf{\Phi}^{\mathcal{V} } \in \mathcal{R}^{N \times K}$       &  		embedding matrix for all nodes				\\
$\mathbf{Z}^{\mathcal{E} } \in \mathcal{R}^{E \times K}$           &  		embedding matrix for all relationships of node-pairs				\\
   
 \bottomrule
 \end{tabular}}\label{tb:notation}
\end{table}

\section{Learning Representations in Semantic Metric Space}
To address the aforementioned problem, we introduce NEMR, which is an unsupervised method and can learn the latent representations of nodes and the relationships between them in metric space.

\subsection{Metric Space}
\label{metric space}
Traditional network representation methods embed nodes into a low-dimensional vector space, and they ofthen use inner-product to measure the similarities of node-pairs, then the relationships for node-pairs can be captured. However, this way can not capture the relationships among neighborhoods of nodes in some cases\cite{hsieh2017collaborative}, an example of which has shown in Fig.3:
\par for a node $v_1$ and its two neighborhoods $u_1$ and $u_2$, two relationships represented as follows:
\begin{align}
&\mathbf{v_1} \cdot \mathbf{u_1} ~= ~ \mathbf{v_1} \cdot \mathbf{u_2}, \nonumber \\
&\mathbf{u_1} \cdot \mathbf{u_2} ~=~\mathbf{0},\label{eq:mistake of relationship}
\end{align}
As neighbors of node $v_1$, nodes $u_1$ and $u_2$ are similar in attributes, but the method based on inner-product can not capture the relationships between $u_1$ and $u_2$ due to they are vertical in the latent space.
\par What's more, the models based on inner-product lack the ability of capturing uncertainty of relationships. Some recent studies have focused on the uncertainty of nodes\cite{bojchevski2017deep,chen2020gaussian} via Gaussian embedding. Similarly, there is uncertainty in the relationship between nodes. In real networks, the uncertainty of the relationship is due to that nodes play different roles, and it is necessary to consider this uncertainty in metric space when modeling the relationship between nodes.
\begin{figure}[!t]
\center
\includegraphics[width=0.7\columnwidth]{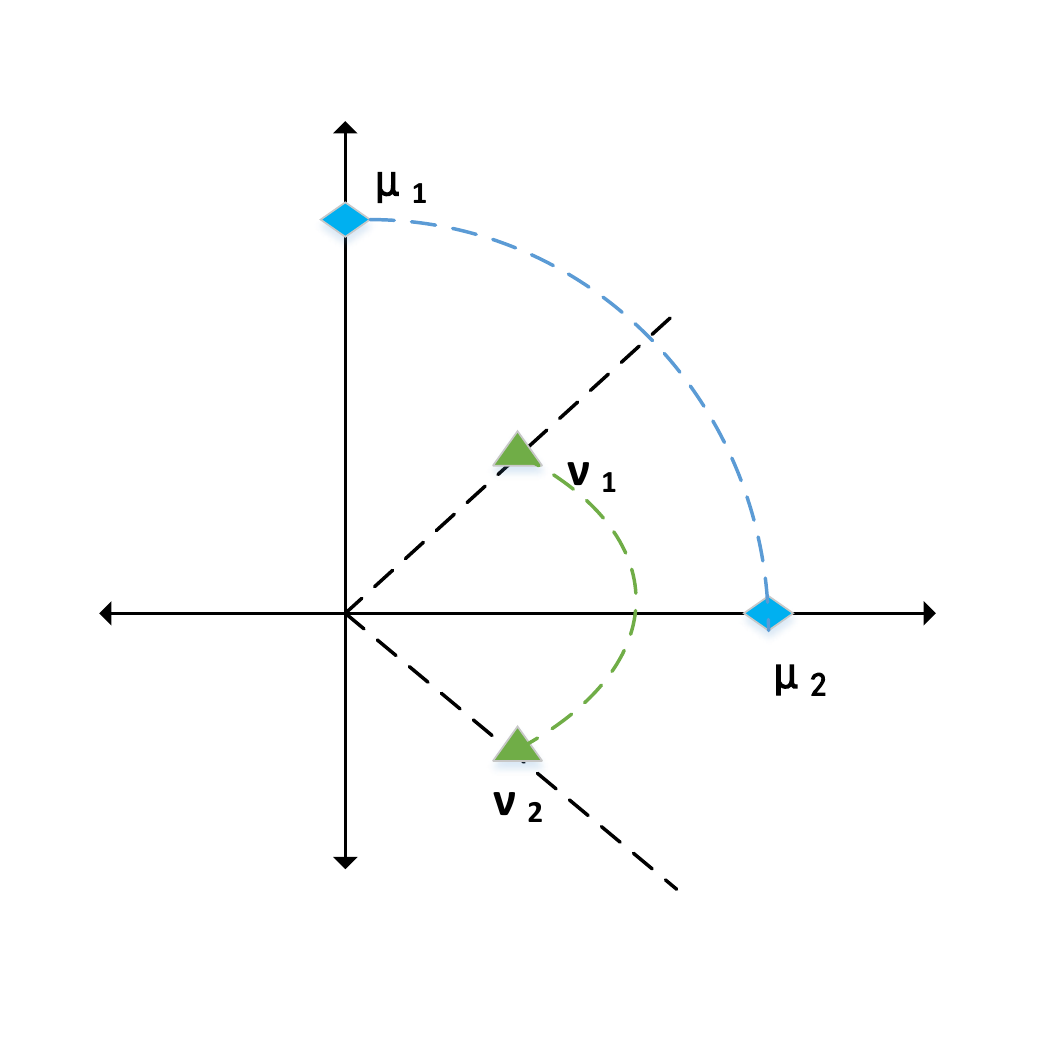}
\caption{\textbf{A case that methods based inner-product can not capture the relationship among neighborhoods. As we can see, nodes $\mu_{1}$ and $\mu_{2}$ are two neighborhoods of node $v_1$, and the relationship between $u_1$ and $u_2$ can not be captured because they are vertical in the latent space.}}
\label{fig:figure2}
\end{figure} 
\par In order to capture the relationships for node-pair well, we model the relationship between two nodes in metric space. For node $v_i$ and $v_j$, we use the function $g(\mathbf{\Phi}_{i}^{\mathcal{V}}, \mathbf{\Phi}_{j}^{\mathcal{V}})$ to represent their relationship, in which $\mathbf{\Phi}_{i}^{\mathcal{V}}$ and $\mathbf{\Phi}_{j}^{\mathcal{V}}$ represent the embedding of node $v_i$ and $v_j$. We propose three implementations of function $g(\cdot)$: 

\par \textbf{2-Norm(2N).} The norm of a vector can simply be understood as its length, or the corresponding distance between two points. Two-normal form is a common way to measure the distance between vectors\cite{dokmanic2015euclidean,song2016deep,zhang2004distance}. Similarly to them, we define the relationship between node $i$ and node $j$ as follow:
\begin{equation}
g(\mathbf{\Phi}_{i}^{\mathcal{V}}, \mathbf{\Phi}_{j}^{\mathcal{V}}) = \left \| \mathbf{\Phi}_{i}^{\mathcal{V}} - \mathbf{\Phi}_{j}^{\mathcal{V}}\right \|_2
\end{equation}
The two norm distance satisfies the critical triangle inequality, which can avoid the case shown in Eq.3.
\par \textbf{Multi-Layer Perceptron (MLP).} A completed MLP consists of three parts: input layer, output layer and hidden layer. MLP can obtain a complex feature transformation through the combination of linear and nonlinear neural networks. In the real world, simple linear function limits the capacity to model the complex relationships between nodes. We use MLP to capture the relationships between nodes, due to its powerful representation ability1:
\begin{equation}
g(\mathbf{\Phi}_{i}^{\mathcal{V}}, \mathbf{\Phi}_{j}^{\mathcal{V}}) = \mathrm{MLP}_{\zeta}(\mathbf{\Phi}_{i}^{\mathcal{V}}  \circ \mathbf{\Phi}_{j}^{\mathcal{V}}) ,
\end{equation}
in which, $\circ$ is the concatenation operation and $\zeta$ is the trainable hyper-parameter of $\mathrm{MLP}$. 

\par \textbf{Variational Inference.} To capture the uncertainty of the relationships of nodes in the semantic metric space, we use a gaussian distribution $\mathcal{N}(\bm{\mu},\bm{\sigma})$ to represent the relationship between nodes, where the mean $\bm{\mu}$ denotes the representation of relationship and the diagonal covariance $\bm{\sigma}$ represents the certainty. We apply an variational inference model to map the $\mathbf{\Phi}_{i}^{\mathcal{V}} - \mathbf{\Phi}_{j}^{\mathcal{V}}$ to the means and variations of the variational posterior distributions 
\begin{equation}
g(\mathbf{\Phi}_{i}^{\mathcal{V}}, \mathbf{\Phi}_{j}^{\mathcal{V}}) = \mathcal{N}(\bm{\mu}_{\phi}(\mathbf{\Phi}_{i}^{\mathcal{V}} - \mathbf{\Phi}_{j}^{\mathcal{V}}), diag(\bm{\Sigma}_{\phi}(\mathbf{\Phi}_{i}^{\mathcal{V}} - \mathbf{\Phi}_{j}^{\mathcal{V}})),
\end{equation}
 where $\phi$ is the hyper-parameter of variational inference model. In what follows, we will describe the inference algorithm in next section.

\subsection{Variational Evidence Lower Bound}
To obtain representations of relationships between nodes in an unsupervised way, the log-likelihood of the observed embeddings can be approximated as follows:
\begin{align}
\log p(\mathbf{\Phi}^{\mathcal{V}}) &= \log \int_{\mathbf{Z}^{\mathcal{E}}} p(\mathbf{\Phi}^{\mathcal{V}}, \mathbf{Z^{\mathcal{E}}}) \, d\mathbf{Z^{\mathbf{\mathcal{E}}}}   \displaybreak[3] \nonumber \\
&=\log \int_{\mathbf{Z}^{\mathcal{E}}} p(\mathbf{\Phi}^{\mathcal{V}}, \mathbf{Z}^{\mathcal{E}}) \frac{q_{\phi}(\mathbf{Z}^{\mathcal{E}} ~|~  \mathbf{\Phi}^{\mathcal{V}})}{q_{\phi}(\mathbf{Z}^{\mathcal{E}} ~|~ \mathbf{\Phi}^{\mathcal{V}})} \, d\mathbf{Z}^{\mathcal{E}} \displaybreak[3]  \nonumber \\
&\ge \mathbb{E}_{q_{\phi}(\mathbf{Z}^{\mathcal{E}}  ~|~ \mathbf{\Phi}^{\mathcal{V}})} \left[   \log \frac{p(\mathbf{\Phi}^{\mathcal{V}}, \mathbf{Z}^{\mathcal{E}})}{q_{\phi}(\mathbf{Z}^{\mathcal{E}} ~|~  \mathbf{\Phi}^{\mathcal{V}})} \right] ,
\label{eq7}
\end{align}
where $q_{\phi}(\mathbf{Z}^{\mathcal{E}} ~|~  \mathbf{\Phi}^{\mathcal{V}})$ is the variational posterior to approximate the true posterior $p(Z^{\mathbf{\mathcal{E}}} ~|~ \mathbf{\Phi}^{\mathcal{V}})$ and 
$q_{\phi}(\mathbf{Z}^{\mathcal{E}} ~|~  \mathbf{\Phi}^{\mathcal{V}})$ is abbreviated as $q_{\phi }$ without causing misunderstanding in the rest of this paper. We use $q_{\phi }$ to approximate the true posterior $p(Z^{\mathbf{\mathcal{E}}} ~|~ \mathbf{\Phi}^{\mathcal{V}})$,  and $\phi$ is the hyper-parameter corresponding to the encoder. 
 For briefly, $q_{\phi}$ can be assumed as a mean-field distribution, which can be represented as follows:
\begin{equation}
q_{\phi}(\mathbf{Z}^{\mathcal{E}} | \mathbf{\Phi}^{\mathcal{V}}) = \prod_{i,j=1}^{N} q_{\phi}(\mathbf{Z}_{i,j}^{\mathcal{E}} ~|~ \mathbf{\Phi}^{\mathcal{V}})
\end{equation}

The joint distribution $p_{\theta}(\mathbf{\Phi}^{\mathcal{V}}, \mathbf{Z}^{\mathcal{E}})$ can be represented as:
\begin{equation}
p(\mathbf{\Phi}^{\mathcal{V}}, \mathbf{Z}^{\mathcal{E}}) = p( \mathbf{Z}^{\mathcal{E}}) \prod_{i,j=1}^{N} p( \mathbf{\Phi}_i^{\mathcal{V}} , \mathbf{\Phi}_j^{\mathcal{V}} | \mathbf{Z}_{i,j}^{\mathcal{E}}),
\end{equation}
where $\theta$ is the hyper-parameter corresponding to the decoder.

By jointing Eq.\ref{eq7}, Eq.8 and Eq.9, the log-likelihood can be rewritted as:

\begin{align}
\log p(\mathbf{\Phi}^{\mathcal{V}}) &\ge \mathbb{E}_{q_{\phi}} \left[ \sum_{i,j=1}^{N} \log p_{\theta}(\mathbf{\Phi}_i^{\mathcal{V}}, \mathbf{\Phi}_j^{\mathcal{V}} ~|~ \mathbf{Z}_{i,j}^{\mathcal{E}}) \right]   \nonumber \\
&- D_{KL}(q_{\phi}(\mathbf{Z}^{\mathcal{E}} ~|~ \mathbf{\Phi}^{\mathcal{V}}) ~\rVert  ~p(\mathbf{Z}^{\mathcal{E}})) \nonumber \\
&= \mathcal{L}(\theta, \phi; \mathbf{\Phi}^{\mathcal{V}})
\end{align}

where $\mathcal{L}(\theta, \phi; \mathbf{\Phi}^{\mathcal{V}})$ is the evidence lower bound (ELBO). $q_{\phi}(\mathbf{Z}^{\mathcal{E}} ~|~ \mathbf{\Phi}^{\mathcal{V}})$ and $p_{\theta}(\mathbf{\Phi}_i^{\mathcal{V}}, \mathbf{\Phi}_j^{\mathcal{V}} ~|~ \mathbf{Z}_{i,j}^{\mathcal{E}})$ are the two parts of VAE, i.e., encoder and decoder, respectively. ~$D_{KL}( \cdot~ \rVert  ~\cdot )$(KL-divergence) is to measure the similarity between variational distribution $q_{\phi}(\mathbf{Z}^{\mathcal{E}} ~|~ \mathbf{\Phi}^{\mathcal{V}})$ and true  posterior distribution \\$p_{\theta}(\mathbf{\Phi}_i^{\mathcal{V}}, \mathbf{\Phi}_j^{\mathcal{V}} ~|~ \mathbf{Z}_{i,j}^{\mathcal{E}})$.
 
\par Following the settings in \cite{kingma2013auto-encoding}, all the priors and posterior distributions are setted to be gaussian distributions. $D_{KL}(q_{\phi}(\mathbf{Z}^{\mathcal{E}} ~|~ \mathbf{\Phi}^{\mathcal{V}}) ~\rVert  ~p(\mathbf{Z}^{\mathcal{E}}))$ can be computed by parameters of gaussian distributions. 
By using the reparameterization trick\cite{kingma2013auto-encoding} and Monte Carlo estimating the expectation terms, we can derivate the ELBO as follow:

\begin{align}
\mathcal{L}(\theta, \phi; \mathbf{\Phi}^{\mathcal{V}}) &= \frac{1}{L} \sum_{i,j=1}^{N} \sum_{l=1}^{L} \log p_{\theta} (\mathbf{\Phi}_i^{\mathcal{V}}, \mathbf{\Phi}_j^{\mathcal{V}} ~|~ \mathbf{Z}_{i,j}^{\mathcal{E}(l)}) \nonumber \nonumber \\
&- D_{KL}(q_{\phi}(\mathbf{Z}^{\mathcal{E}} ~|~ \mathbf{\Phi}^{\mathcal{V}}) ~\rVert  ~p(\mathbf{Z}^{\mathcal{E}})), 
\end{align}
where 
\begin{align}
\mathbf{Z}_{i,j}^{\mathcal{E}(l)} = \bm{\mu}_{(\mathbf{\Phi}_i^{\mathcal{V}} - \mathbf{\Phi}_j^{\mathcal{V}})} + \bm{\sigma}_{({\mathbf{\Phi}_i^{\mathcal{V}} - \mathbf{\Phi}_j^{\mathcal{V}})}} \odot \bm{\epsilon}^{l},  \bm{\epsilon}^{l} \backsim \mathcal{N}(0,\mathbf{I}).
\end{align}

%
As shown in Eq.11, by optimizing the lower bound we can obtain the latent embedding $\mathbf{Z}_{i,j}^{\mathcal{E}}$, which can be regarded as the relationship between node $i$ and $j$.

\subsection{Learning}
In order to solve the problem that the methods based on inner-product can not capture the complex relationships between nodes well, we adopt three methods to model the relationship between nodes in metric space(see in \S\ref{metric space}). In what follows, we propose two constrains of path(see in Definition.4 and Definition.6): the equivalence of multiple-paths and the natural order of a single-path, and our loss function can be divided into two independent parts corresponding to them.
\par The first part can be represented as follows:
\begin{equation}
\mathcal{L}_{mul} = \sum_{i,j=1}^{N} ~\sum_{P_1,P_2 \in \bm{\mathcal{S}}_{i,j}} {\rm G}(R_{P_1}, ~R_{P_2}),
\end{equation}
that is, locally per node-pair $i$ and $j$ we want the sum of the relationships of each path to be the same. $\bm{\mathcal{S}}_{i,j}$ represent the set of path between node $i$ and $j$, $P_1$ and $P_2$ are two paths in $\bm{\mathcal{S}}_{i,j}$. $R_{P}$ represent the sum of relationships of path $P$, assume that $P= \{v_1,v_2, ...... v_{t-1}, v_t\}$, the $R_p$ can be calculated as follows:
\begin{equation}
R_P=g(\mathbf{\Phi}_{1}^{\mathcal{V}}, \mathbf{\Phi}_{2}^{\mathcal{V}})+ ...... +g(\mathbf{\Phi}_{t-1}^{\mathcal{V}}, \mathbf{\Phi}_{t}^{\mathcal{V}}),
\end{equation}
in which, function $g(\cdot)$ (detailed in \S\ref{metric space}) can map the relationship between nodes into metric space. If we map the relationship into metric space by 2-Normal(2N)/Multilayer Perceptron(MLP), the $g(\mathbf{\Phi}_{i}^{\mathcal{V}}, \mathbf{\Phi}_{j}^{\mathcal{V}})$ is a constant/vector, $R_P$ can be calculated by  element-wise addition. Specifically, if we map the relationship into metric space by variational Inference(VI), the $g(\mathbf{\Phi}_{i}^{\mathcal{V}}, \mathbf{\Phi}_{j}^{\mathcal{V}})$ is a gaussian distribution. Following \cite{he2015learning}, the sum of two gaussian distributions is also a gaussian distribution, suppose $\mathcal{P}_1 \sim \mathcal{N}(\bm{\mu}_1, \bm{\Sigma}_1)$, $\mathcal{P}_2 \sim \mathcal{N}(\bm{\mu}_2, \bm{\Sigma}_2)$, $\mathcal{P}_{1+2} \sim \mathcal{N}(\bm{\mu}_1 + \bm{\mu}_1, \bm{\Sigma}_2 + \bm{\Sigma}_2)$.

\par Accordingly, the function $\rm{G}(\cdot)$ can be written in the following forms:

\begin{itemize}

\item For 2N and MLP, we have:
\begin{align}
{\rm G} (R_{P_1}, R_{P_2})= (R_{P_1} - R_{P_2})^{2} .
\end{align}

\item For VI, we have:
\begin{align}
~~~~{\rm G} (R_{P_1}, R_{P_2}) = D_{KL}(R_{P_1} ~\rVert~ R_{P_2})^{2}, 
\end{align}
where $D_{KL}(\cdot)$ is the KL-divergence to measure the similarity of two distributions, which can be represented as follow:
\begin{align}
&D_{KL}(\mathcal{N}_{2} \rVert \mathcal{N}_{1}) =\frac{1}{2}[tr(\Sigma_{1}^{-1}\Sigma_{2}) \\
&+(\mu_{1} - \mu_{2})^T\Sigma_{i}^{-1}(\mu_{1} - \mu_{2})  \nonumber - L - log\frac{det(\Sigma_{2})}{det(\Sigma_{1})}] .
\end{align}
\end{itemize}

\par The second part of our loss function can be represented as follows:
\begin{equation}
\mathcal{L}_{sin} = -\sum_{P \in \bm{\mathcal{T}}} ~ \sum_{i,j \in P}  \exp (r_{i,j}^{\prime} - R_{i,j}),
\end{equation}
in which, $\bm{\mathcal{T}}$ is the set of single-paths in network, $r_{i,j}^{\prime}$ represents the direct distance in relation metric space, $R_{i,j}$ is the sum of relationships between node $i$ and $j$.  The $\exp (r_{i,j}^{\prime} - R_{i,j})$ can be derived as Eq.15, Eq.16 and Eq.17.

\par Finally, we optimize the following joint objective function that is the sum corresponding to Eq.13 and Eq.18:
\begin{equation}
\mathcal{L} = \lambda\mathcal{L}_{mul} + (1-\lambda)\mathcal{L}_{sin},
\end{equation}

in which, $\lambda$ is a trade off parameter to balance the two parts of the loss function $\mathcal{L}$. We can optimize the parameters such that the loss $\mathcal{L}$ is minimized, the two constrains we proposed can be satisfied. When the model has converged, we can get high quality node embeddings. Especially, for the variational inference method, we need to add the ELBO to the objective function for optimization, and the loss function can be rewritten as follows:
\begin{equation}
\mathcal{L} = \lambda\mathcal{L}_{mul} + (1-\lambda)\mathcal{L}_{sin} + \mathcal{L}(\theta, \phi; \mathbf{\Phi}^{\mathcal{V}}).
\end{equation}

The overview of NEMR is shown in Algorithm 1.

\begin{algorithm}[!t]
\small
  \DontPrintSemicolon
  \LinesNumbered
  \SetKwInOut{Input}{Input}\SetKwInOut{Output}{Output}
  \SetKw{Or}{or}
  \Input
  {
  	the set of nodes: $\bm{\mathcal{V}}$,  the set of edges: $\bm{\mathcal{E}}$, adjacency matrix: \textbf{A},
	the set of multi-paths:   $\bm{\mathcal{S}}$,
	the set of single-path:   $\bm{\mathcal{T}}$,
	embedding size: $K$,
	the metric function: $g(\cdot)$	 ~ in ~\S\ref{metric space}\\	
	
  }
  \Output
  {
  	node embeddings: $\bm{\Phi}^{\bm{\mathcal{V}}}$ \\
  }
  {
  \textbf{foreach}  $ i,j \in \bm{\mathcal{V}} $ ~ \textbf{do}\\ 	
  	~~~~~~~~$\bm{\Phi}_{i}^{\bm{\mathcal{V}}}$ $\leftarrow$ Uniform($\frac{-1}{\sqrt{K}}$)\\
	~~~~~~~~$\bm{\Phi}_{j}^{\bm{\mathcal{V}}}$ $\leftarrow$ Uniform($\frac{-1}{\sqrt{K}}$)\\
	~~~~~~~~$\bm{Z}_{i,j}^{\bm{\mathcal{E}}}$ $\leftarrow$ $g(\bm{\Phi}_{i}^{\bm{\mathcal{V}}}, \bm{\Phi}_{i}^{\bm{\mathcal{V}}})$\\
  }
  \textbf{loop}\\
  {
  	~~~~\textbf{foreach}  $\bm{\mathcal{S}}_{i,j}  \in \bm{\mathcal{S}} $ \\
	~~~~~~~~ \textbf{foreach}  $P_1, P_2  \in \bm{\mathcal{S}_{i,j}} $ ~ \textbf{do}\\
  	~~~~~~~~~~~~ obtain $R_{P_1}$ and $R_{P_2}$ by Eq.14 										\\
	~~~~~~~~~~~~ obtain ${\rm{G}}(R_{P_1}, R_{P_2})$ by Eq.15 or Eq.16						\\
	~~~~~~~~~~~~ update embeddings w.r.t. $\mathcal{L}$\\
	~~~~~~~~\textbf{end for}					\\
	~~~~\textbf{end for}						\\
	~~~~\textbf{foreach}  $P  \in \bm{\mathcal{T}} $ \\	
	~~~~~~~~ \textbf{foreach} $i,j \in P$  ~ \textbf{do}\\
	~~~~~~~~~~~~ obtain $\exp (r_{i,j}^{\prime} - R_{i,j})$ \\
	~~~~~~~~~~~~ update embeddings w.r.t. $\mathcal{L}$\\
	~~~~~~~~\textbf{end for}					\\
	~~~~\textbf{end for}\\
  }
  \textbf{end loop}
  \caption{THE LEARNING ALGORITHM OF NEMR.}
\label{alg:overview}
\end{algorithm}

\section{EXPERIMENTS}

\setlength{\tabcolsep}{3mm}{
\begin{table*}[t]
\center
\caption{Link prediction performance with embedding size L = 128.}
\resizebox{0.9\textwidth}{!}{
\begin{tabular}{ccccccccccc}
\toprule
method     & \multicolumn{2}{c}{Cora} & \multicolumn{2}{c}{DBLP} & \multicolumn{2}{c}{BlogCatalog} & \multicolumn{2}{c}{Flickr} & \multicolumn{2}{c}{Pubmed}        \\
\cmidrule{1-11}
                   & AUC          & AP           & AUC         & AP         & AUC           & AP           & AUC              & AP      &AUC      &AP         \\
DeepWalk   &.734   &.721         &.716           &.722        &.693            &.706          &.711               &.691	      &.732	       &.716               \\
node2vec   & 74.5        & .729        & .725       & .756      & .740         & .714        & .742            & .762               &.719        &.724                 \\
TADW       & .836        & .814        & .763       & .784      & .776         & .771        & .673            & .634                 &.847	&.830     						\\
TRIDNR   & .861        & .895        & .805       & .813      & .805         & .816        & .804  		& .812 		&.819	&.822				\\
GAE        & .967       & .961       & .941       & .937      & .816         & .810        & .828           & .836 	&.916	&.925					\\
\hline
NEMS\_2N   & .941        & .943        & .935      &.946       &.801          & .803        & .824           & .832  	&.947 	&.945\\
NEMS\_MLP  & .924        & .911       & .924       &.930     & .812         & .829        & .904            & .913  	&.939	&.952\\
NEMS\_VL  & \textbf{.973}      & \textbf{.979}       & \textbf{.984}       & \textbf{.981}      & \textbf{.834}         & \textbf{.840}        & \textbf{.926}    & \textbf{.918}    &\textbf{.971}    &\textbf{.979}\\
\bottomrule           
\end{tabular}}
\end{table*}}

To verify the effectiveness of our proposed algorithm, we compared NEMS with existing state-of-the-art algorithms in the tasks of link prediction and node classification. And we also analyze the impact of the parameter sensitivity of our algorithm, e.g., the embedding size and the percentage of training edges.
\subsection{Data Sets}
\label{datasets}
We use several public graph datasets, the statistics of which is provided in Tab. 2:

\begin{itemize}

\item \textbf{Cora}\cite{bojchevski2018deep}:The Cora datasets is citation networks, which is composed of a large number of academic articles. Nodes are publications and edges are citation links.

\item \textbf{DBLP}\cite{pan2016tri}: DBLP dataset \footnote{Available from: \url{http://dblp.uni-trier.de/xml/}.}, which consists of bibliography data in computer science. Each paper may cite or be cited by other papers, which naturally forms a citation network.

\item \textbf{BlogCatalog}\cite{meng2019co}: This dataset is a social relationship network, which is crawled from the BlogCatalog website  \footnote{Available from: \url{http://www.blogcatalog.com/}.}. BlogCatalog is composed of the bloggers and their social relationships. The labels of nodes indicate the interests of the bloggers .

\item \textbf{Flickr} \cite{meng2019co}: Flickr is a social network where users can share pictures and videos. \footnote{Available from: \url{http://www.Flickr.com/}.} In this dataset, each node is a user and each side is a friend relationship between users. In addition, each node has a label that identifies the user’s interest group.
 
\item \textbf{Pubmed}\cite{bojchevski2018deep}: Pubmed is a public search database that provides biomedical paper and abstract search service. \footnote{Available from: \url{http://pubmed.com.cutestat.com/}.} In this dataset, nodes are publications and edges are citation links.

\end{itemize}

\begin{table}
\center 
\caption{\textbf{Statistics of the data sets used in the experiments.}}
\resizebox{0.9\columnwidth}{!}{
\begin{tabular}{@{}l@{~~}*{4}{c@{~~}}c@{}}
\toprule
& \textbf{\#Nodes} & \textbf{\#Edges} &\textbf{\#Attributes} & \textbf{\#Labels}  \\
\cmidrule{2-5}
	 \textbf{Cora}  & 2,708 & 5,429 & 1,433 & 7  \\
         \textbf{DBLP} & 17,716 & 105,734 & 1,639 & 4  \\
         \textbf{BlogCatalog} & 5,196 & 171,743 & 8,189 & 6  \\
         \textbf{Flickr} & 7,575 & 239,738 & 12,047 & 9 \\  
         \textbf{Pubmed}  &19,717 &44,338 &500 &3     \\
         \bottomrule
\end{tabular}}
\end{table}

\subsection{Baslines}
To evaluate the performance of our model, we compare our NEMR with the following baselines in this paper:

\begin{itemize}
\item \textbf{DeepWalk\cite{perozzi2014deepwalk}:}This method is inspired by Skip-Gram model, and it adopts random walk to extract node sequences. 	By doing so, Skip-Gram model can be used to obtain the node embeddings.
\item \textbf{Node2Vec\cite{grover2016node2vec}:} A variant of DeepWalk, which apply Depth-First Search(DFS) and Breadth-First Search(BFS) into random walk. Node2Vec adopts a flexible neighborhood sampling strategy to capture the Hojophily and Homogeny of the network.
\item \textbf{TADW\cite{yang2015network}:}  A method based on low-rank matrix factorization which considers graph structure and text features.
\item \textbf{GAE\cite{kipf2016variational}:} The GAE model  is an example of VAEs that learns representations for attributed networks.
\item \textbf{TRIDNR\cite{pan2016tri}:} TRIDNR is an extension of DeepWalk\cite{perozzi2014deepwalk} model. This model considers network structures, the contexts of nodes and the labels of nodes when obtaining node embeddings. 
\end{itemize}

We implement all the baselines by the codes released by the authors. For DeepWalk model and its variants, the window size $w$ is set to 4 and the size of negative sample $N$ is set to 5, the length of random walk $t$ is set to 20. For all baselines, the embedding size $K$ is set to 128. The initial learning rate of Adam is set as 0.001. For our model, when extracting the path between nodes $i$ and $j$, we filter out the path which length is greater than 10(Cora) or 20(DBLP, BlogCatalog and Flickr). For our inference neural network, we use a 128-dimensional hidden layer and 128-dimensional latent variables in all experiments. 

\begin{figure}[!t]
\center
\includegraphics[width=1.0\columnwidth]{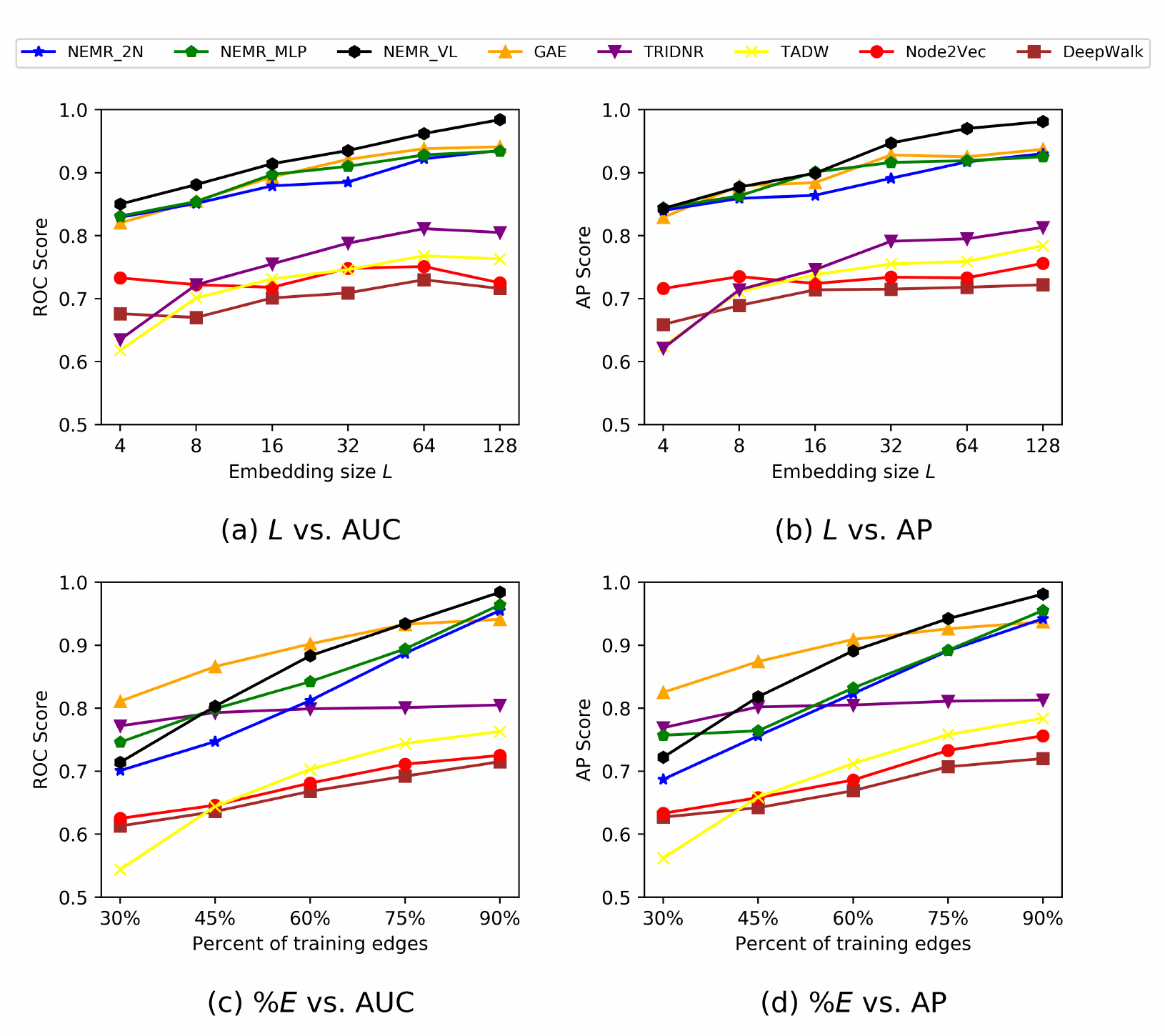}
\vspace{-1.9em}
\caption{Link prediction performance with different embedding sizes and percentages of training edges on DBLP dataset. Compared with other baselines, NEMR still perform well under low dimension of embedding size and small size of training edges.}
\label{fig:figure}
\end{figure} 

\begin{figure*}[!t]
\center
\includegraphics[width=1.95\columnwidth]{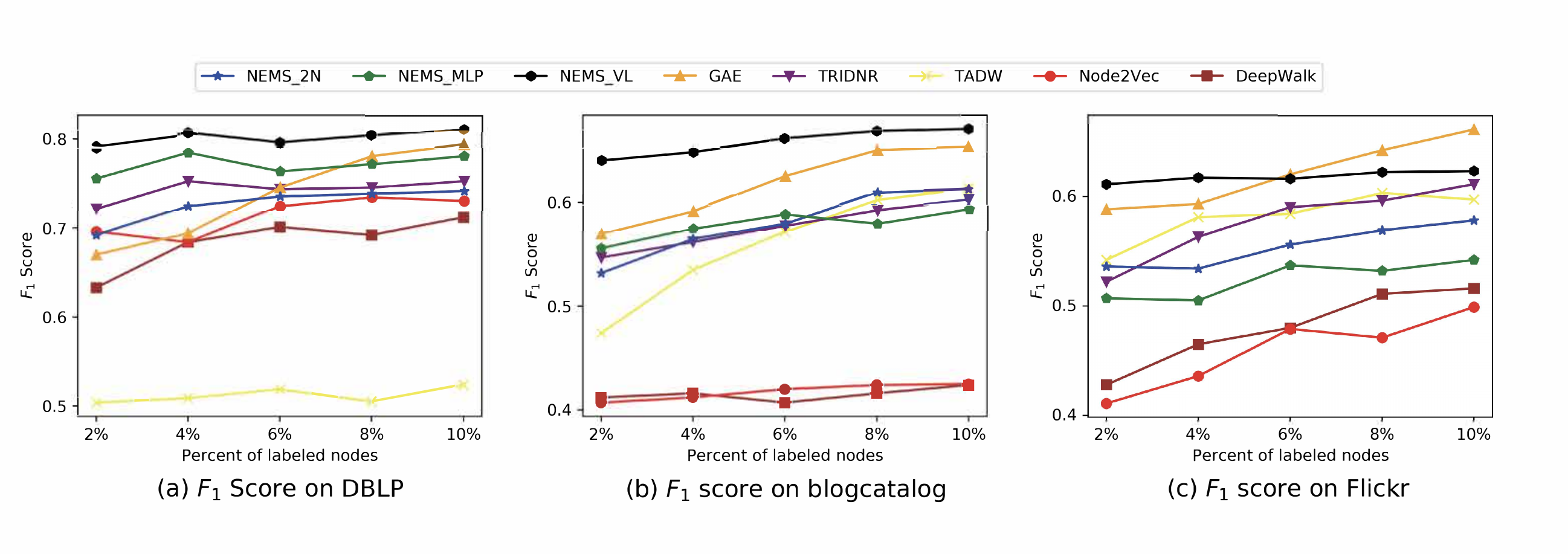}
\vspace{-2em}
\caption{Node classification performance for different  percentages of training edges. Our NEMR model achieved the best results on all DBLP and BlogCatalog.}
\label{fig:figure5}
\end{figure*} 
\subsection{Link Prediction}
Link prediction aims to predict the future interactions of nodes in the network, and it can be also used in multiple tasks, e.g., the recommendations of friends and products. In this subsection, we compare our NEMR with the baselines on link prediction task. Same to \cite{kipf2016variational,wang2016structural}, we create a validation/test set that contains 5\%/10\% randomly selected edges respectively and equal number of randomly selected non-edges. To measure the performance of link prediction task, we report the area under the ROC curve (AUC) and the average precision (AP) scores for each method.

\par Tab. 3 shows the link prediction performance of our \\NEMR and the baselines on the five datasets  mentioned in \S\ref{datasets}. We can see that, our NEMR model significantly outperforms the baselines across all datasets. In DBLP, BlogCatalog and Filckr datasets, NEMR\_VL all achieves the best performance, that is because our NEMR\_VL considers the equivalence of multiple-paths which is sensitive for relation-rich datasets. What's more, NEMR\_VL maps the relationship between nodes into a distribution to capture the uncertainty which can better learn network structures. NEMR\_2N and NEMR\_MLP also show excellent performance which demonstrate that modelling the relationship in metric space is effective to learn better node embeddings.

\subsection{Sensitivity analysis.}
The dimension of the embedding is sensitive, if the dimension of embedding is too small, the relationship between nodes can not be captured well, otherwise it will suffer from overfitting and increasing computational complexity. In this section, We analyze the effect of embedding dimension on performance, when $L$ is equal to $4$, $8$, $16$, $32$, $64$, $128$. Finally, we evaluate the performance w.r.t. the percentage of training edges varying from $30\%$ to $85\%$, averaged over 10 trials.

As we can see In Fig.4(a) and 4(b), our model still shows good performance under low embedding sizes. Even for \\$L$~=~4, our NEMR\_VL achieves the best performance compared to the other baselines. That is because NEMR embed the relationship between nodes as a gaussian distribution which allow us to capture uncertainty of the relationship between nodes.

\par We also evaluate NEMR's sensitivity about the size of the training edges, and the percentage of training edges is set from $15\%$ to $85\%$, repeated 10 times.  As we can see in Fig.4(c) and Fig.4(d), with the increase of the percentage of training edges, the performance of our model improves significantly. Our NEMR model obtains the network structure through two constrains(see in Definition. 4 and Definition. 6) of the paths, the increasing size of training edges means that NEMR can capture the relationships between nodes better.

\begingroup
\setlength{\tabcolsep}{3.0mm}{
\begin{table}
\center
\caption{Node classification performance.}
\resizebox{0.9\columnwidth}{!}{
\begin{tabular}{ccccccccc}
\toprule
method      & \multicolumn{1}{c}{DBLP} & \multicolumn{1}{c}{BlogCatalog} & \multicolumn{1}{c}{Flickr}         \\
\cmidrule{1-4}
DeepWalk   &.712   &.424         &.516               		\\
node2vec   & .731        & .425        & .499                      \\
TADW       & .524       & .614       & .597                       \\
TRIDNR   & .753        & .603        & .611        \\
GAE        & .794        & .654        & \textbf{.661}           \\
\hline
NEMS\_2N   & .742        & .613       & .578            \\
NEMS\_MLP  & .781        & .594       & .542            \\
NEMS\_VL  & \textbf{.811}       & \textbf{.671}       & .623            \\
\bottomrule           
\end{tabular}}
\end{table}}
\endgroup
\vspace{1em}

\subsection{Node Classification}
Node classification is also one of the important tasks to detect the performance of embedding models.
In this section, we perform node classification task to evaluate the performance of the learned embeddings and compare with the baseline methods. We choose three datasets(DBLP,BlogCatalog and Flickr) which have ground-truth classes. To be specific, We first sample a small number of nodes as training data and the rest is test data. Same to \cite{perozzi2014deepwalk}, we use one-vs-rest logistic regression for node classification, and the training data size is $10\%$, the results are reported in Tab 4. To obtain the effect of the size of the training data on the classification result, we make a series of choices on the percentage of the training data($2\%$, $4\%$, $6\%$, $8\%$,$10\%$). This process is repeated for 10 times, and we use the average score of F1-score as evaluation metrics. The results are reported in figure 5.

As we can see in Tab.4, our proposed NEMR\_VL performs the best in DBLP and BlogCatalog datasets, and shows competitive performance in other datasets compared with other network embedding baselines. Again, it proves that the proposed model can better capture the network structures. Fig.5 shows the influence of the proportion of data used for training logistic regression on experimental performance. NEMR\_VL model can obtain the optimal performance under various conditions which proves the stability of our model. 

\section{CONCLUSION}
We proposed a novel Network Embedding on Metric of Relation algorithm, NEMR, an unsupervised approach that  can efficiently learn the embeddings of both nodes and their relationships for large scale networks. 
In metric space, we model the relationship between nodes in three ways, where the variational inference embeds the relationship as a gaussian distribution allowing us to capture uncertainty. 
Furthermore, we proposed the equivalence between multiple paths which can obtain the rich information of the edges in network. In our experiments, we have evaluated the performance of NEMR and the baseline algorithms on seven real-world social networks. The experimental results have demonstrated that our NEMR model is effective for learning node embeddings.  As future work we consider to incorporate attributes of nodes into the model and apply our NEMR into a series of networks, e.g.,  tree network and star network.

\vspace{1em}
\par \textbf{ACKNOWLEDGMENTS} This work is supported by National Key R \& D Program of China Project \\ 
\#2017YFB0203201, Australian Research Council Discovery Project DP150104871.

\bibliographystyle{elsarticle-num}
\bibliography{ref}

\begin{thebibliography}{10}
\expandafter\ifx\csname url\endcsname\relax
  \def\url#1{\texttt{#1}}\fi
\expandafter\ifx\csname urlprefix\endcsname\relax\def\urlprefix{URL }\fi
\expandafter\ifx\csname href\endcsname\relax
  \def\href#1#2{#2} \def\path#1{#1}\fi

\bibitem{li2018streaming}
J.~Li, K.~Cheng, L.~Wu, H.~Liu, Streaming link prediction on dynamic attributed
  networks (2018) 369--377.

\bibitem{li2017attributed}
J.~Li, H.~Dani, X.~Hu, J.~Tang, Y.~Chang, H.~Liu, Attributed network embedding
  for learning in a dynamic environment, in: Proceedings of the 2017 ACM on
  Conference on Information and Knowledge Management, ACM, 2017, pp. 387--396.

\bibitem{perozzi2014deepwalk}
B.~Perozzi, R.~Al-Rfou, S.~Skiena, Deepwalk: On learning of social
  representations, in: Proceedings of the 20th ACM SIGKDD international
  conference on Knowledge discovery and data mining, ACM, 2014, pp. 701--710.

\bibitem{wang2017community}
X.~Wang, P.~Cui, J.~Wang, J.~Pei, W.~Zhu, S.~Yang, Community preserving network
  embedding (2017) 203--209.

\bibitem{zhang2017user}
D.~Zhang, J.~Yin, X.~Zhu, C.~Zhang, User profile preserving social network
  embedding, in: IJCAI International Joint Conference on Artificial
  Intelligence, 2017.

\bibitem{tang2015pte}
J.~Tang, M.~Qu, Q.~Mei, Pte: Predictive text embedding through large-scale
  heterogeneous text networks, in: Proceedings of the 21th ACM SIGKDD
  International Conference on Knowledge Discovery and Data Mining, ACM, 2015,
  pp. 1165--1174.

\bibitem{huang2017label}
X.~Huang, J.~Li, X.~Hu, Label informed attributed network embedding (2017)
  731--739.

\bibitem{cao2015grarep}
S.~Cao, W.~Lu, Q.~Xu, Grarep: Learning graph representations with global
  structural information, in: Proceedings of the 24th ACM international on
  conference on information and knowledge management, ACM, 2015, pp. 891--900.

\bibitem{mikolov2013efficient}
T.~Mikolov, K.~Chen, G.~Corrado, J.~Dean, Efficient estimation of word
  representations in vector space, arXiv preprint arXiv:1301.3781 (2013).

\bibitem{grover2016node2vec}
A.~Grover, J.~Leskovec, node2vec: Scalable feature learning for networks, in:
  Proceedings of the 22nd ACM SIGKDD international conference on Knowledge
  discovery and data mining, ACM, 2016, pp. 855--864.

\bibitem{tang2015line}
J.~Tang, M.~Qu, M.~Wang, M.~Zhang, J.~Yan, Q.~Mei, Line: Large-scale
  information network embedding, in: Proceedings of the 24th international
  conference on world wide web, International World Wide Web Conferences
  Steering Committee, 2015, pp. 1067--1077.

\bibitem{chen2017pne:}
W.~Chen, X.~Mao, X.~Li, Y.~Zhang, X.~Li, Pne: Label embedding enhanced network
  embedding (2017) 547--560.

\bibitem{liu2018content}
J.~Liu, Z.~He, L.~Wei, Y.~Huang, Content to node: Self-translation network
  embedding (2018) 1794--1802.

\bibitem{kipf2017semi-supervised}
T.~Kipf, M.~Welling, Semi-supervised classification with graph convolutional
  networks (2017).

\bibitem{huang2017accelerated}
X.~Huang, J.~Li, X.~Hu, Accelerated attributed network embedding. (2017)
  633--641.

\bibitem{hsieh2017collaborative}
C.-K. Hsieh, L.~Yang, Y.~Cui, T.-Y. Lin, S.~Belongie, D.~Estrin, Collaborative
  metric learning, in: Proceedings of the 26th international conference on
  world wide web, International World Wide Web Conferences Steering Committee,
  2017, pp. 193--201.

\bibitem{pan2016tri}
S.~Pan, J.~Wu, X.~Zhu, C.~Zhang, Y.~Wang, Tri-party deep network
  representation, Network 11~(9) (2016) 12.

\bibitem{qiu2018network}
J.~Qiu, Y.~Dong, H.~Ma, J.~Li, K.~Wang, J.~Tang, Network embedding as matrix
  factorization: Unifying deepwalk, line, pte, and node2vec, in: Proceedings of
  the Eleventh ACM International Conference on Web Search and Data Mining,
  2018, pp. 459--467.

\bibitem{xing2002distance}
E.~P. Xing, M.~I. Jordan, S.~Russell, A.~Y. Ng, Distance metric learning with
  application to clustering with side-information (2002) 521--528.

\bibitem{qian2015fine}
Q.~Qian, R.~Jin, S.~Zhu, Y.~Lin, Fine-grained visual categorization via
  multi-stage metric learning, in: Proceedings of the IEEE Conference on
  Computer Vision and Pattern Recognition, 2015, pp. 3716--3724.

\bibitem{lebanon2006metric}
G.~Lebanon, Metric learning for text documents, IEEE Transactions on Pattern
  Analysis and Machine Intelligence 28~(4) (2006) 497--508.

\bibitem{wang2005learning}
D.~Wang, X.~Ma, Y.~Kim, Learning pseudo metric for intelligent multimedia data
  classification and retrieval., Journal of Intelligent Manufacturing 16~(6)
  (2005) 575--586.

\bibitem{cheng2019network}
X.~Cheng, L.~Ji, R.~Huang, R.~Cui, Network embedding with deep metric learning,
  IEICE Transactions on Information and Systems~(3) (2019) 568--578.

\bibitem{chen2018pme:}
H.~Chen, H.~Yin, W.~Wang, H.~Wang, Q.~V.~H. Nguyen, X.~Li, Pme: Projected
  metric embedding on heterogeneous networks for link prediction (2018)
  1177--1186.

\bibitem{bourlard1988auto-association}
H.~Bourlard, Y.~Kamp, Auto-association by multilayer perceptrons and singular
  value decomposition, Biological Cybernetics 59~(4) (1988) 291--294.

\bibitem{kingma2013auto-encoding}
D.~P. Kingma, M.~Welling, Auto-encoding variational bayes, arXiv: Machine
  Learning (2013).

\bibitem{kingma2015adam:}
D.~P. Kingma, J.~Ba, Adam: A method for stochastic optimization (2015).

\bibitem{jiang2016variational}
Z.~Jiang, Y.~Zheng, H.~Tan, B.~Tang, H.~Zhou, Variational deep embedding: An
  unsupervised and generative approach to clustering, arXiv: Computer Vision
  and Pattern Recognition (2016).

\bibitem{kingma2014semi-supervised}
D.~P. Kingma, S.~Mohamed, D.~J. Rezende, M.~Welling, Semi-supervised learning
  with deep generative models 27 (2014) 3581--3589.

\bibitem{kipf2016variational}
T.~Kipf, M.~Welling, Variational graph auto-encoders, arXiv: Machine Learning
  (2016).

\bibitem{creswell2019denoising}
A.~Creswell, A.~A. Bharath, Denoising adversarial autoencoders, IEEE
  Transactions on Neural Networks 30~(4) (2019) 968--984.

\bibitem{dosovitskiy2016generating}
A.~Dosovitskiy, T.~Brox, Generating images with perceptual similarity metrics
  based on deep networks, arXiv: Learning (2016).

\bibitem{yang2017fast}
C.~Yang, M.~Sun, Z.~Liu, C.~Tu, Fast network embedding enhancement via high
  order proximity approximation., in: IJCAI, 2017, pp. 3894--3900.

\bibitem{yang2015network}
C.~Yang, Z.~Liu, D.~Zhao, M.~Sun, E.~Chang, Network representation learning
  with rich text information, in: Twenty-Fourth International Joint Conference
  on Artificial Intelligence, 2015.

\bibitem{wang2016structural}
D.~Wang, P.~Cui, W.~Zhu, Structural deep network embedding, in: Proceedings of
  the 22nd ACM SIGKDD international conference on Knowledge discovery and data
  mining, ACM, 2016, pp. 1225--1234.

\bibitem{liu2017semantic}
Z.~Liu, V.~W. Zheng, Z.~Zhao, F.~Zhu, K.~C.-C. Chang, M.~Wu, J.~Ying, Semantic
  proximity search on heterogeneous graph by proximity embedding, in:
  Thirty-First AAAI Conference on Artificial Intelligence, 2017.

\bibitem{hamilton2017inductive}
W.~L. Hamilton, Z.~Ying, J.~Leskovec, Inductive representation learning on
  large graphs (2017) 1024--1034.

\bibitem{bordes2013translating}
A.~Bordes, N.~Usunier, A.~Garciaduran, J.~Weston, O.~Yakhnenko, Translating
  embeddings for modeling multi-relational data (2013) 2787--2795.

\bibitem{wang2014knowledge}
Z.~Wang, J.~Zhang, J.~Feng, Z.~Chen, Knowledge graph embedding by translating
  on hyperplanes, in: Twenty-Eighth AAAI conference on artificial intelligence,
  2014.

\bibitem{vilnis2014word}
L.~Vilnis, A.~McCallum, Word representations via gaussian embedding, arXiv
  preprint arXiv:1412.6623 (2014).

\bibitem{bojchevski2017deep}
A.~Bojchevski, S.~Gunnemann, Deep gaussian embedding of graphs: Unsupervised
  inductive learning via ranking, arXiv: Machine Learning (2017).

\bibitem{chen2020gaussian}
Y.~Chen, J.~Pu, X.~Liu, X.~Zhang, Gaussian mixture embedding of multiple node
  roles in networks, World Wide Web 23~(2) (2020) 927--950.

\bibitem{dokmanic2015euclidean}
I.~Dokmanic, R.~Parhizkar, J.~Ranieri, M.~Vetterli, Euclidean distance
  matrices: Essential theory, algorithms, and applications, IEEE Signal
  Processing Magazine 32~(6) (2015) 12--30.

\bibitem{song2016deep}
H.~O. Song, Y.~Xiang, S.~Jegelka, S.~Savarese, Deep metric learning via lifted
  structured feature embedding (2016) 4004--4012.

\bibitem{zhang2004distance}
Y.~G. Zhang, C.~Zhang, D.~Zhang, Distance metric learning by knowledge
  embedding, Pattern Recognition 37~(1) (2004) 161--163.

\bibitem{he2015learning}
S.~He, K.~Liu, G.~Ji, J.~Zhao, Learning to represent knowledge graphs with
  gaussian embedding (2015) 623--632.

\bibitem{bojchevski2018deep}
A.~Bojchevski, S.~Gunnemann, Deep gaussian embedding of graphs: Unsupervised
  inductive learning via ranking (2018).

\bibitem{meng2019co}
Z.~Meng, S.~Liang, H.~Bao, X.~Zhang, Co-embedding attributed networks, in:
  Proceedings of the Twelfth ACM International Conference on Web Search and
  Data Mining, ACM, 2019, pp. 393--401.

\end{thebibliography}

\end{document}